\documentstyle[aps,psfig,epsfig,multicol]{revtex}
\begin{document}
\def\be{\begin{equation}}
\def\ee{\end{equation}}
\def\bc{\begin{center}}
\def\ec{\end{center}}
\def\bea{\begin{eqnarray}}
\def\eea{\end{eqnarray}}
\newcommand{\bleq}{\ifpreprintsty
                   \else
                   \end{multicols}\vspace*{-3.5ex}\widetext{\tiny
                   \noindent\begin{tabular}[t]{c|}
                   \parbox{0.493\hsize}{~} \\ \hline \end{tabular}}
                   \fi}
\newcommand{\eleq}{\ifpreprintsty
                   \else
                   {\tiny\hspace*{\fill}\begin{tabular}[t]{|c}\hline
                    \parbox{0.49\hsize}{~} \\
                    \end{tabular}}\vspace*{-2.5ex}\begin{multicols}{2}
                    \narrowtext \fi}

\title{Size of quantum networks}
\author{Ginestra Bianconi}
\address{D\'epartement de Physique Th\'eorique, Universit\'e de Fribourg
P\'erolles,CH-1700, Switzerland}
\maketitle
\begin{abstract}
The metric structure of bosonic scale-free networks
 and  fermionic Cayley-tree  networks is analyzed focousing on the
directed distance of nodes from the origin. The topology of the netwoks strongly depends on the dynamical parameter $T$, called temperature. At $T=\infty$ we show analytically that the two networks have a similar behavior: the distance of a generic node from the origin of the network scales as the logarithm of the number of nodes in the network.
At $T=0$ the two networks have an opposite behavior: 
the bosonic network remains very clusterized (the distance from the origin remains constant as the network increases the number of nodes) while  the fermionic network grows following a single branch of the tree and  the distance from the origin  goes as a power-law of the number of nodes in the network.
\end{abstract}
\begin{multicols}{2}


Complex networks representing systems of interacting units have been studied and classified \cite{Stanley_rev,Kim_rev} according to  their different geometrical and topological properties. In between complex networks scale-free networks,with power-law connectivity distribution, have been found to describe different systems of nature and society\cite{RMP,DoroRev,Strogatz}.
Recently several models \cite{RMP,DoroRev,Strogatz,FitnessG} have been
    formulated that generate such structures, the prototype of them being the BA model\cite{BA}.
Scale-free networks are particulary interesting because their  highly inhomogeneous structure  induces peculiar effects in the dynamical models that can be defined
on them, as the absence of  percolation \cite{Havlin} and  epidemic threshold\cite{Vespignani},the   infinite Curie temperature for the paramagnetic to ferromagnetic transition in the frame of the Ising model\cite{Stauffer_ising,Doro_ising,Leone_ising,G_ising} and the good associative memory of the Hopfiled model
 defined on network with large average connectivity \cite{Hopfield}.
Besides, the investigation of the metric stucture \cite{NatureWWW,Newman_metric,Doro_metric,Wuchty,ultrasmall}
  of these networks    is of great interest. 
It was first empirically found \cite{NatureWWW} and then analytically derived \cite{Newman_metric,Doro_metric} that
scale-free networks are characterized by having a mean distance $<d>$ between nodes scaling like the logarithm of the system size $N$, $<d>\sim\log(N)$.

The similarities between the structure of a BA network and a traditional Cayley tree have been recently studied. It has been found that they share high similarities. In fact they are both generated by the subsequent addition of a same elementary unit (a node connected to $m$ links) attached to the rest of the network  in the direct or in the reversed direction \cite{misto}. The symmetry  between these two types of networks  is  evident if we assume that each node $i$ has an innate quality or ``energy'' $\epsilon_i$ and that the dynamics is parametrized by a variable $T$ called temperature that introduce a ``thermal noise''
as it has been done
 in self-organized models \cite{Vergeles1,Temp_G,Vergeles2}.
It is possible, then to observe that the BA model becomes a limiting case of a scale-free network described by a Bose distribution of the energies to which the incoming links point \cite{bose}, while the growing Cayley tree network is described by a Fermi distribution of the energies at the interface\cite{fermi}.
Quantum networks (the bosonic scale-free  network and the fermionic Cayley tree network) evolve around a well defined core of initial nodes.

In this paper we focus  our attention on the metric structure of quantum networks and in particular on their size, i.e. we extimated  the distance measured over directed paths of a generic node $i$ from the origin of the network and  its dependence on the time $t_i$ in which the node $i$ has been added  to the network.
We derive the expression for the average  value of the distance $<\ell(t_i)>$    of node $i$ introduced at time $t_i$ from the origin,  mesuread on directed paths, in quantum networks at    $T=\infty$.
We find in agreement with \cite{NatureWWW,Newman_metric,Doro_metric} that $<\ell(t_i)>\sim \log(t_i)$ in the bosonic scale-free network and we show for $T=\infty$ a similar behavior in fermionic networks.

At different values of $T$ the topology of the network change drastically. For energy distribution functions $p(\epsilon)\rightarrow 0$ for $\epsilon \rightarrow 0$ the exponent $\gamma$ of the power-law connectivity distribution $P(k)\sim k^{-\gamma}$ goes from $\gamma=3$ at $T=\infty$ to $\gamma=2$ for $T=0$ and there is a phase transition at a cirtical temperature $T_c$ below which a finite fraction of all the links is connected to a single node.
Assuming that the bosonic network is a reasonable model for growing scale-free networks,we will have, for example, that the citation network \cite{Redner} with $\gamma\sim 3$ would correspond to a $T=\infty$ dynamics while the incoming component of the World-Wide-Web with $\gamma_{in}=2.1$\cite{BA,Broder} would correspond to a low temperature dynamics.
As $T$ decreses also  the behavior of $<\ell(t_i)>$ changes and
the distance of node $i$ from  the origin depends less strongly on the time $t_i$ of its arrival. At sufficiently low temperature $\ell(t_i)$ remains constant as a function of $t_i$.
On the contrary in the fermionic network, at $T=0$, when the dynamics becomes extremal, the network evolves far away from the origin and the distance of a node $i$ from the origin of the network grows as a power-law of the time $t_i$ of its arrival in the network.

\subsection{Distances from the origin in the bosonic network}

The bosonic network is a generalization of the well known BA network\cite{BA}. 
In this model, a new node with $m$ links is added to the network at each time step. 
Each node $i$ has an innate quality or 'energy' $\epsilon_i$ extracted from a probability  distribution $p(\epsilon)$. The way the new links are attached  follows a generalized preferential attachment rule:  the probability $\Pi_j$ that an existing node $j$ acquires 
a new link  depends both on  its connectivity $k_j(t)$ and  its energy $\epsilon_j$, i.e.
\be
\Pi_j =  \frac{e^{-\beta \epsilon_j}k_j(t)}{\sum_s e^{-\beta \epsilon_s}k_s(t)}.\label{pref}
\ee
with the parameter $\beta=1/T$ tuning the relevance of the energy $\epsilon_j$ with respect to the connectivity $k_j(t)$.
This network displays a power law connectivity distribution $P(k) \sim k^{-\gamma}$ with $\gamma \in [2,3]$ depending on the $p(\epsilon)$ distribution and the inverse temperature $\beta$ \cite{bose}.
In the $T=\infty$ limit  ($\beta=0$) the network reduces to a scale-free BA network \cite{BA}
with an average connectivity $k_i$ of node $i$ that grows in time as a power-law with exponent $1/2$
\be
k_i(t)=m\sqrt{\frac{t}{t_i}},
\label{ki}
\ee
where $t_i$ is the time in which node $i$ was added to the network.
 The probability $p_{i,j}$ that two nodes $i$ and $j$ are connected by a link can be calculated from Eq. ($\ref{pref}$). At $T=\infty$  ($\beta=0$), taking into account that at each time $m$ new links are added to the network, the probability $p_{i,j}    $ is given by $m$ times  Eq. ($\ref{pref}$). After substituting $(\ref{ki})$ into Eq. $(\ref{pref})$ we obtain,
\be
p_{  i,  j}=\frac{m}{2}\frac{1}{\sqrt{t_i t_j}}.
\label{p_ij}
\ee

The number of directed paths of length $\ell$ connecting  node $i$, introduced   in the network at time
$t_{i}$,  to  node $i_0$ belonging to  the original core  of the network  ($t_{i_0}=1$) is given by the mean value of the number of paths  connecting node
$i$  to $i_0$ and passing through the points $i_0,i_1,
i_2,\dots i_{\ell-1},i_{\ell}=i$ with $t_{n+1}>t_n$. Indicating each
node with the time of its arrival in the network,   the probability of any directed path is given by the product $\Pi_{n=1}^{\ell} p_{n-1,n}$ with $t_{n-1}<t_n$. In order to find $n_{\ell}(t_i)$ we should sum over all possible paths.
  Replacing the sum
over the nodes with the integrals over $t_n$, we obtain for
$n_\ell(t_i)$,

\be
n_{\ell}(t_i)= \int_{1}^{t_{i}} dt_1
\int_{t_1}^{t_{i}} dt_2\dots \int_{t_{\ell-2}}^{t_{i}}dt_{\ell-1}p_{     0 ,  1}p_{  1,  2}\dots p_{{\ell-1},\ell} .
\ee

Using $t_0=1$ and Eq.~$(\ref{p_ij})$ valid in the $T=\infty$ limit, we obtain
\bea n_{\ell}(t_i)&=&{\left(\frac{m}{2}\right)}^{\ell}\int_{1      }^{t_{i}}
dt_1\int_{t_1}^{t_{i}} dt_2\dots \nonumber \\
& & \dots \int_{t_{\ell-2}}^{t_{i}}dt_{\ell-1}
\frac{1}{t_1}\frac{1}{t_2}\dots\frac{1}{t_{\ell-1}}
\frac{1}{\sqrt{       t_{i}}} \nonumber \\ &=&
\frac{1}{(\ell-1)!}{\left(\frac{m}{2}\log\left(     {t_{i}}
         \right)\right)}^{\ell-1}\frac{m}{2\sqrt{       t_{i}}}. \eea

This means that the mean distance between node $i$ and a node $i_0$, such that $t_{i_0}=1$, calculated only on directed paths, follows a Poisson
distribution with average size
\be
<\ell(t_i)>=\frac{m}{2}\log(t_{i}). \label{elle_mean}\ee

\begin{figure}
\centerline{\epsfxsize=2.5in \epsfbox{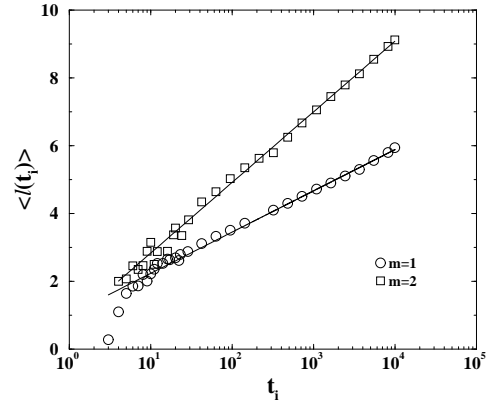}} \caption{Mean distance from the origin  $<\ell (t_i)>$ of the nodes arrived  at time $t_i$ in a BA network (a bosonic network at $T=\infty$) with $m=1,2$. The solid lines indicate the theoretical prediction Eq. $(\ref{elle_mean}$).} \label{p_elle.fig}
\end{figure}

The distribution of the number of directed paths of length $\ell$ starting from the origin of the network is  proportional to the integral of $n_{\ell}(t_i)$ over $t_i$ 
\be
P(\ell)\propto\int_1^t n_{\ell}(t') dt'= (-m)^{\ell} \left(1-\frac{\Gamma(\ell,\log(t)/2)}{\Gamma({\ell})}\right).
\label{distri.eq}
\ee
In Fig. $\ref{distri.fig}$ we show the agreement between the numerical results and Eq. $(\ref{distri.eq})$ for a bosonic network of $10^4$ nodes at $T=\infty$  and $m=1,2$.

\begin{figure}
\centerline{\epsfxsize=2.5in \epsfbox{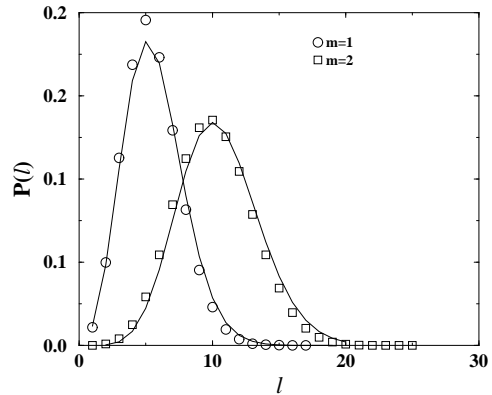}} \caption{Distribution of the number of directed paths of length $\ell$ in a BA network (a bosonic network at $T=\infty$) for $m=1,2$. The solid lines are the analytic predictions Eq.~$(\ref{distri.eq})$.} \label{distri.fig}
\end{figure}

The   topology of the bosonic network changes as a function of the  temperature $T=1/\beta$.
In particular for a  distribution $p(\epsilon)$ such that $p(\epsilon)\rightarrow0$ as $\epsilon\rightarrow 0$ we know\cite{bose} that there is a critical temperature $T_c$  below which the
network has a topological transition and its structure is dominated by a single node that grabs a finite fraction of all the links. For $T>T_c$ ($\beta<\beta_c$) the network is in the so-called ``fit-gets-rich'' phase (FGR phase)  while for $T<T_c$ ($\beta>\beta_c$) the network is in the so-called Bose-Einstein condensate phase (BE phase).
To visualize this transition we have plotted the  bosonic  network with 
an energy distribution 
\be
p(\epsilon)=\frac{1}{\theta+1} \epsilon^{\theta}, \mbox{        } \epsilon\in(0,1)
\label{pe.eq}
\ee
where $\theta=0.5$, $m=1$, above (Fig. $\ref{bosonic.fig}$a) and below (Fig. $\ref{bosonic.fig}$b) the phase transition.
\begin{figure}
\centerline{\epsfxsize=2.5in \epsfbox{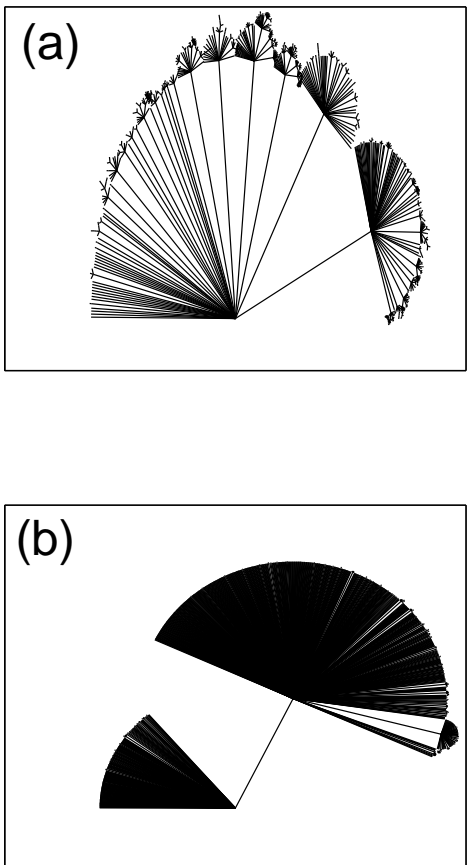}} \caption{Graphic representation of the bosonic network. Graph (a) represents a network with $m=1$, energy distribution given by Eq. $(\ref{pe.eq})$ where $\theta=0.5$ at $T=\infty$ ($\beta=0$ -the BA network) and  graph (b) represents a network with the same parameters $m$ and $\theta$ but with  $T=1/\beta=0.33$ (network in the Bose-Einstein condensate phase). The number of nodes   both networks is $N=10^3$. } \label{bosonic.fig}
\end{figure}
The network has been designed in order to underline the hierarchical structure of the network. Starting from the  single node at the origin of the tree, we have placed all the nodes that are directly attached to it on a semicircle of unitary radius, each node $i$ separated from the next one by  the angle $\Delta \alpha_i$ proportional to its connectivity, i.e. \be
\Delta \alpha_i=\frac{k_i}{\sum_{j\in N(i)}k_j}\ee where $N(i)$ are the nearest neighbors of node $i$ added at a time $t_j>t_i$. We have repeated the same construction for all the nodes of the network  in such a way that all the nearest neighbors of node $i$  are on a semicircle of  radius $r_i$ with
\be
r_i=r_k \frac{k_i}{\sum_{j\in N(i)}k_j},
\ee
where $k$ is the node to which the node $i$ has been attached at time $t_i$.
From Fig. $\ref{bosonic.fig}$ it  is clear the change in the topology of the network at the critical temperature with the emergence of a single node that grabs a finite fraction of all the links in the Bose-Einstein condensate phase.

As the topology of the network changes, the behavior of $\ell(i)$ as a function of $t_i$ changes too. 
In fact  we have
\be
<\ell(i)>=a(\beta) \log(t_i)
\label{a_beta.eq}
\ee
where the coefficient $a(\beta)$  is a decreasing function of the inverse temperature $\beta=1/T$.
\begin{figure}
\centerline{\epsfxsize=2.5in \epsfbox{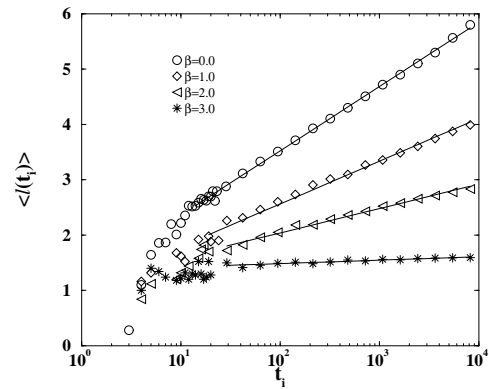}} \caption{Distances form the origin in a bosonic network with $m=2$ and $\theta=0.5$ as a function of $\beta$.} \label{p_elle_beta.fig}
\end{figure}
In Fig. $\ref{p_elle_beta.fig}$ we report $<\ell(i)>$ for a bosonic network with $p(\epsilon)$ of the type $(\ref{pe.eq})$ with $\theta=0.5$ and $m=2$ at different values of the inverse temperature $\beta$, above and below the critical value $\beta_c=1.7$ \cite{bose}.
 In   oredr to illustrate the changement of the topology of the network above and below the critical temperature in Fig. $\ref{trans.fig}$ we report the behavior of different relevant structural quantities  for  a network of size $N=10^4$,  $\theta=0.5$ and with $m=2$ around the critical inverse temperature temperature $\beta_c=1.7$ . We report the  fraction of links attached to the most connected node, the exponent of the power-law component of the connectivity distribution, the clustering coefficient and the coefficient $a(\beta)$.
The fraction of links attached to the most connected network  $k_{max}(\beta)/N$ is the order parameter of the FGR-BE phase transition and  increases as a  function of  $\beta$. The data reported in Fig.~$\ref{trans.fig}$ are mediated over $100$ runs. The connectivity distribution of the bosonic network contains a power-law component plus a pick indicating the condensation phenomena that appears for $\beta>\beta_c$. In Fig.~ $\ref{trans.fig}$ we report  the exponent $\gamma(\beta)$ of the power-law component of the connectivity distribution that decreases as a function of $\beta$ with an asymptotic value of $\gamma=2$. The data reported on Fig.~$\ref{trans.fig}$  are mediated over $100$ runs. The clustering coefficient $C(\beta)$ increases at the transition point while the coefficient $a(\beta)$ slowly decreases saturating toward a zero value for $\beta>3.0$.
The data reported in Fig.~$\ref{trans.fig}$ for these last  two quantities data are mediated over $10$ run.

\begin{figure}
\centerline{\epsfxsize=3.5in \epsfbox{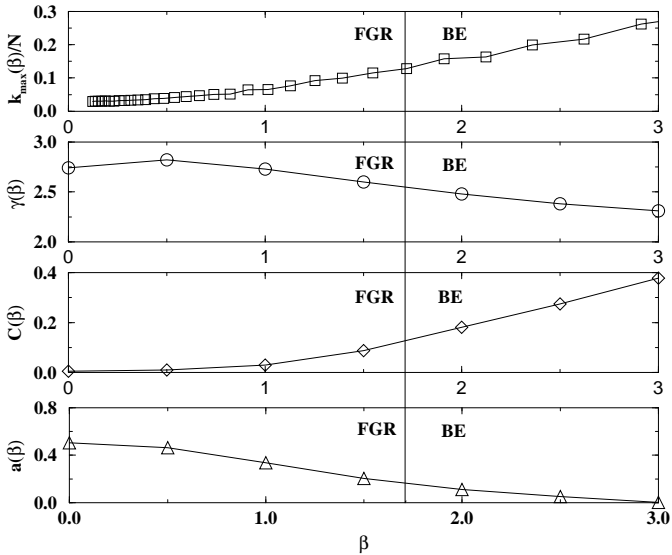}} \caption{Relevant structural quantities in a bosonic network of size $N=10^4$ with $m=2$ and $\theta=0.5$ as a function of the inverse temperature $\beta$.} \label{trans.fig}
\end{figure}

\subsection{Distances from the origin in the fermionic network}

The fermionic network  \cite{fermi} is a growing Cayley tree, where the innate qualities of the nodes (energies) define their  different branching tendency.
Starting at time  $t=1$ from a node $i_0$ at the origin of the network, the node $i_0$  at time  $t=2$ grows and $m$ new nodes are directly connected to it.
Each node $i$ has an energy $\epsilon_i$ extracted from a given $p(\epsilon)$ distribution. 
At each timestep  a  new  node with connectivity one (at the interface) is chosen to branch, giving rise to $m$
new nodes.

We assume that  nodes with higher energy  are more likely to grow than
lower energy ones. In particular we take $\Pi_i$, the probability  that a  
node $i$ of the interface (with energy $\epsilon_i$)  grows at time $t$, to be 
\be
\Pi_i=\frac{e^{\beta \epsilon_i}}{\sum_{j\in {\rm Int(t)}} e^{\beta
\epsilon_j}}. \label{TreeEn.pi1} \ee where the sum  in the
denominator is  extended to all  nodes $j$ at the
interface $Int(t)$ at time $t$. The model  depends on the inverse temperature
 $\beta=1/T$.  In the $\beta\rightarrow 0$ limit, high and low energy
 nodes are  equally probable to grow and the model reduces to the {\it Eden model}. 
In the $\beta\rightarrow \infty$ limit
the dynamics becomes extremal and only the nodes with the
highest energy value are allowed to grow. In this case the model reduces to
{\it invasion percolation }\cite{IP,Vanderwalle} on a Cayley tree.

 Let us assume that
  $\beta=0$ ($T=\infty$).
  The probability $\Pi_i$
that a node $i$ of the interface $Int(t)$
 grows at time $t$ is given by
\be
\Pi_{i}=\frac{1}{N_{Int}(t)}, \label{TreeEden.pi}\ee where
$N_{Int}(t)$ is the total number of active nodes. Since at each
timestep a node of the interface branches, 
and $m$ new  nodes  are generated, after $t$ timesteps the
model generates an interface of $N_{Int}(t)$ nodes, with
 \be
N_{Int}(t)=(m-1)t+1. \label{Interface.eq} \ee

 Lets denote by $\rho(t,t_i)$ the probability that a node,
 born at time $t_i$ is still at the interface at time $t$.
Since every node $i$ of the network grows with probability $\Pi_i$ 
Eq. $(\ref{TreeEden.pi})$ if it is at the interface, in mean
field $\rho(t,t_i)$ follows
\be
\frac{\partial \rho(t,t_i)}{\partial t}= -\frac{ \rho(t,t_i)}
{N_{Int}(t)}. \label{TreeEden.eq}\ee Replacing
$(\ref{Interface.eq})$ in $(\ref{TreeEden.eq})$  in the limit
$t\rightarrow \infty$ we get the solution
\be
\rho(t,t_i)={\left(\frac{t_i}{t}\right)}^{1/(m-1)}.
\label{TreeEden.dyn} \ee Consequently each node $i$ that  arrives
at the interface at time $t_i$, remains at the interface with a
probability that decreases  in time as a power-law.
The probability $p_{i,j}$ that a node $i$ is attached to a node $j$ (arrived in the network at a later time $t_j>t_i$) is given by the r. h. s. of Eq. $(\ref{TreeEden.pi})$ calculated at time $t_j$. Taking into account Eq.$(\ref{TreeEden.dyn})$ and the rate $m$ of the addition of new nodes, we obtain for $p_{  i,  j}$
\be
p_{  i,  j}=\frac{m}{(m-1)t_j} \left({\frac{t_i}{t_j}}\right)^{1/(m-1)}.
\label{p_ij_tree}
\ee

 The number 
$n_{\ell}(t_i)$ of paths of length $\ell$ that connect a node $i$,  introduced at time
$t_{i}$, to the origin $i_0$  is given by the average number of paths connecting a node 
 $i$ to a node $i_0$ and passing through the points $i_0,i_1
i_2,\dots i_{\ell-1},i_{\ell}=i$ with $t_{n+1}>t_n$. Indicating each
node with the time of its arrival in the network and the sum
over the nodes with the integrals over $t_n$, we obtain for
$n_{\ell}(t_i)$,

 \be
n_{\ell}(t_i)= \int_{1      }^{t_{i}} dt_1
\int_{t_1}^{t_{i}} dt_2\dots
\int_{t_{\ell-2}}^{t_{i}}dt_{\ell-1} p_{     0    1}p_{  1,  2}\dots p_{{\ell-1},\ell } \ee
 and,
using $(\ref{p_ij_tree})$  we obtain, with a calculation analog to 
$ (5)$,   

\bea
 n_\ell(t_{i})
\frac{1}{(\ell-1)!}{\left(\frac{m}{m-1}\log\left(     {t_{i}}
         \right)\right)}^{\ell-1}\left({\frac{1      }{t_{i}}}\right)^{1/(m-1)} . 
\eea

\begin{figure}
\centerline{\epsfxsize=2.5in \epsfbox{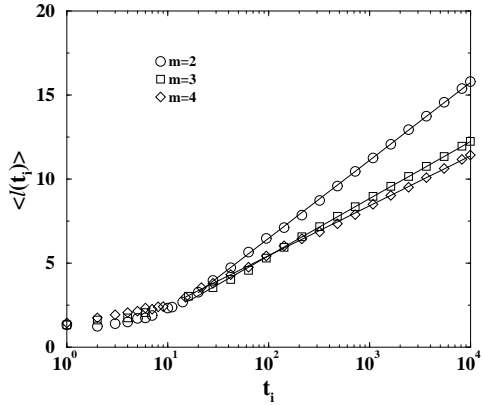}} \caption{Distance from the origin in a fermionic network at $T=\infty$ ($\beta=0$) for networks of $N=10^4$ nodes with $p(\epsilon)$ uniform between zero and one and with $m=2,3,4$. The solid lines are the theoretical predictions Eq.($\ref{elle_tree}$). } \label{tree_elle.fig}
\end{figure}
This means that the mean distance between a node $t_{i}$ and
the origin  follows a Poisson
distribution with average size
\be
<\ell(t_i)>=\frac{m}{m-1}\log(t_{i}). \label{elle_tree}\ee
As in the bosonic network, in the fermionic network at infinite temperature
($\beta=0$) the distance $<\ell(t_i)>$ of node $i$ from the origin grows logarithmically with the time $t_i$.

In Fig. $\ref{tree_elle.fig}$ we report the analytical simulation of a fermionic network  with $p(\epsilon)$ uniform between zero and one, at $T=\infty$ and $m=2,3,4$.

As the temperature decreases, the topology of the network changes drastically, in Fig. $\ref{fermionic.fig}$ we show the Cayley tree with $p(\epsilon)=1$, $\epsilon\in(0,1)$  and $m=2$ at infinite temperature ($\beta=0.0$) and at low temperature ($\beta=20.0$). At high temperature the network grows homogeneously in each direction while at low temperature it evolves following only a single branch of the tree.
\begin{figure}
\centerline{\epsfxsize=2.5in \epsfbox{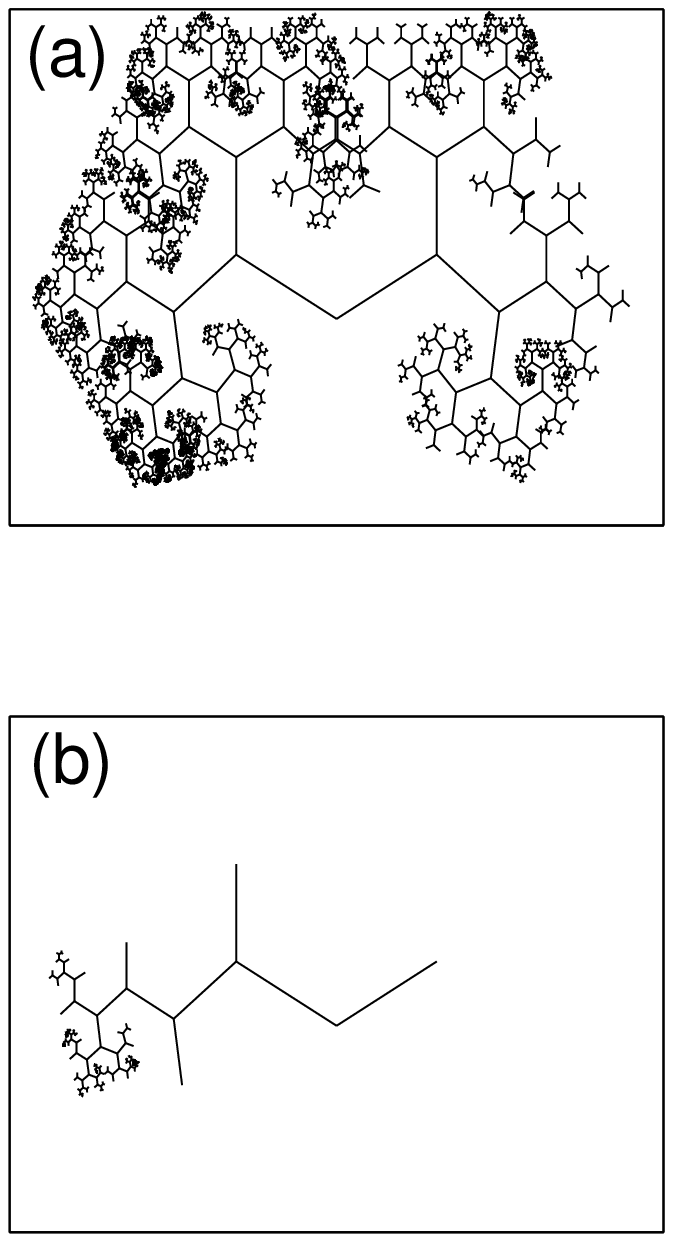}} \caption{The fermionic network with $m=2$ at $T=\infty$ (Graph (a)) and at temperature $T=0.05$ (Graph (b)).The number of nodes in both networks are $N=10^4$. } \label{fermionic.fig}
\end{figure}

The distance of a node $i$ from the origin of the networks grows  logarithmically  with $t_i$ at $T=\infty$ ($\beta=0$). As the temperature decreases  the behavior of $<\ell(t_i)>$  gets steeper. For  $p(\epsilon)$  uniformly distributed between zero and one, in the extremal case $T=0$ ($\beta=\infty$) when the node of highest energy grows deterministically at each time step, we have a dramatic change in the behavior  and $<\ell (i)>$  grows as a power-law of $t_i$,
\be
<\ell (i)>\propto (t_i)^{\zeta}
\label{p-l.eq}
\ee
 $\zeta=0.55\pm 0.05$ from the numerical results reported in Fig. $\ref{tree_elle_beta.fig}$.

In conclusion we have shown that bosonic and fermionic network are not only simmetricaly built\cite{misto} but also at $T=\infty$ they are characterized by a distance $<\ell(t_i)>$ from the origin that grows like the logarithm  of the time $t_i$. On the contrary, in the limit $T=0$ they behave in a opposite way: the bosonic network stays highly clusterized with a distance from the origin that remains constant as the network evolves, in the fermionic network the distance $<\ell(t_i)>$  grows like a power-law of the time $t_i$.

\begin{figure}
\centerline{\epsfxsize=2.5in \epsfbox{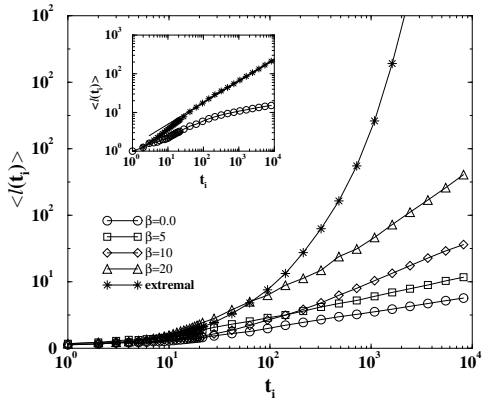}} \caption{Distance from the origin in a femionic network with $m=2$ and $p(\epsilon)$ uniformly distributed between zero and one at different temperatures. At $\beta=0$ ($T=\infty$) we have the predicted logarithmic behavior Eq. $(\ref{elle_tree})$ while in the extremal case $\beta=\infty$ ($T=0$) the network grows as a power-law of the network size. The solid line in the Inset is the power-law fit Eq. $(\ref{p-l.eq})$ with $\zeta=0.55\pm0.05$. } \label{tree_elle_beta.fig}
\end{figure}
\section{Acknowledgments}
 We are grateful to A. Capocci, P. Laureti and  Y.-C. Zhang  for useful comments and discussions.
This paper has been financially supported by the Swiss National Foundation,
under Grant No.2051-067733.02/1 and by the European Commission - Fet Open projectCOSIN IST-2001-33555.


\end{multicols}
\end{document}